# In Bad Faith: Assessing Discussion Quality on Social Media

Celia Chen[1][0009-0008-9967-8968], Alex Leitch[1][0009-0009-0298-7260], William Jordan Conway[1], Eric Cotugno[1], Emily Klein[2], Rajesh Kumar Gnanasekaran[1], Kristin Buckstad Hamilton[1], Casi Sherman[1], Celia Sterrn[1], Logan C. Stevens[1], Rebecca Zarrella[1], Jennifer Golbeck[1][0000-0003-3684-307X]

[1]University of Maryland, College Park MD 20740, USA
[2]University of Albany, Albany, NY 12227, USA
`jgolbeck@umd.edu`

**Abstract.** The quality of a user's social media experience is determined both by the content they see and by the quality of the conversation and interaction around it. In this paper, we look at replies to tweets from mainstream media outlets and official government agencies and assess if they are good faith, engaging honestly and constructively with the original post, or bad faith, attacking the author or derailing the conversation. We assess automated approaches that may help in making this determination and then show that within our dataset of replies to mainstream media outlets and government agencies, bad faith interactions constitute 68.3% of all replies we studied, suggesting potential concerns about the quality of discourse in these specific conversational contexts. This is particularly true from verified accounts, where 91.7% of replies were bad faith. Given that verified accounts are algorithmically amplified, we discuss the implications of our work for understanding the user experience on social media.

## 1 Introduction

Social media is an important source of news and information, with half of Americans getting news from social platforms [1]. The sources, accuracy, and amplification of social media content is important, as is the conversation around that content – the thing that makes it social. Online cultures can be engaged, curious, and organized around sharing high-quality and high-value information, or they can be unhealthy, riddled with harassment, bad faith engagement, and hate speech.

In this paper, we propose a definition and set of criteria for differentiating good and bad faith interactions on social media as a way of measuring one aspect of an online environment's conversational health. Using a codebook built on those criteria, we code a set of replies on popular tweets from mainstream media outlets and official US government accounts. This dataset forms the basis for addressing two main research questions:

RQ1: Can automated techniques accurately assess whether a social media reply is in good or bad faith?
RQ2: What is the state of good vs. bad faith interactions on Twitter/X
    RQ2a: How does verified status relate to good and bad faith interaction on Twitter/X?

For RQ1, we analyze if LLMs, particularly ChatGPT, can read a tweet and reply and accurately label that reply as good or bad faith. Our results show for the dataset studied here, ChatGPT performs at the same level as a human coder, suggesting it may be appropriate for automatically labeling data at scale.

We use both the human-labeled data and a larger set annotated by ChatGPT to further assess the alignment between human and automated coding, and also to assess the state of discourse on Twitter/X. Our results suggest the vast majority of replies on the types of posts we studied are in bad faith and that replies from Twitter/X's Verified accounts engage in bad faith at a significantly higher rate than unverified accounts.

We emphasize that our findings are specifically applicable to the context of direct replies to high-profile mainstream media and government accounts with substantial engagement (over 100 comments). While this represents a window into public discourse with authoritative information sources, we acknowledge the limitations in generalizing these findings to the entirety of Twitter/X conversations or to other social media platforms.

## 2    Related Work and Methodology

### 2.1    Related Work

Previous work has examined conversation quality through various lenses, including natural language processing of online comments [2] and network analysis of how incivility spreads within social graphs [3]. Platform features and design choices influence discussion quality, with research showing that certain network affordances can improve discussion quality [4], while the effectiveness of moderation techniques varies across platforms [5]. Network effects also play a role, with studies finding that small groups of users can influence overall discourse quality [6], creating patterns that affect how views spread over time [7].

### 2.2    Theoretical Foundations and Opereational Definition

Drawing on philosophical work from Sartre [8] and Johannesen [9] and recent frameworks from Roberts-Miller [10] and Craig [11], we developed criteria for distinguishing good and bad faith interactions. Bad faith engagement is characterized by "cynical consciousness" and willful denial of truth [8], while good faith embodies authenticity, self-awareness, and commitment to seeking understanding [9].

We acknowledge that classifying interactions into a binary framework represents a simplification of communicative behaviors that exist along a spectrum. Our coding framework focuses on identifying predominant characteristics to determine whether interactions primarily advance or hinder meaningful discourse.

**Bad faith** comments demonstrate: dismissal of evidence without substantive engagement; strategic derailment through topic shifting; pseudo-engagement that prevents meaningful dialogue; personal attacks and ad hominem arguments; deliberate misrepresentation of facts; inflammatory or derogatory language; and speculative accusations without evidence.

**Good faith** comments exhibit: direct engagement with the original topic; evidence-based reasoning; constructive disagreement or criticism; genuine questions seeking clarification; recognition of issue complexity; respectful tone; and willingness to consider alternative viewpoints.

### 2.3 Data Collection

Twitter/X was selected as a site for this study as it offers a robust collection of fixed, short-format texts in English. This baseline offers good generalizability for posts based in anglophone countries, and mobility to non-X platforms with comparable content formatting, such as Threads or BlueSky.

We collected tweets from mainstream media accounts (NBC News, CBS News, CNN, Washington Post, New York Times, Wall Street Journal, Associated Press, Reuters) based on their presence in Pew studies [12], and official US government accounts from a federal directory. For each account, we used Twitter/X's advanced search to select all 2024 posts with at least 100 comments, ensuring strong engagement within the established Twitter/X ecosystem.

In total, 601 posts met our criteria: 441 from mainstream media (led by NBC News with 198 posts, CBS News with 121) and 160 from 26 government agencies (led by State Department with 50 posts, NASA with 21). We collected 52,469 total replies, of which 31,283 were unique.

Twitter/X's Terms of Use prohibit the public release of this data, but researchers who want access can email jgolbeck@umd.edu to request a copy.

### 2.4 Coding Process

We randomly selected 400 tweet-reply pairs for human coding. Two coders independently labeled each reply, with a third coder resolving disagreements through majority vote. This established our ground truth dataset.

After human coding, ChatGPT-4 independently labeled each reply using a prompt incorporating our codebook criteria:

"I'm going to ask you whether a reply to a tweet is a good faith engagement. Here are the characteristics of good vs. bad faith engagements:

Characteristics of bad faith comments: Dismissal of data – when factual data is presented and ignored or waved off; Generalization – broad unsupported statements;

Lack of engagement of non-constructive engagement with the original post – conflating issues, derailing the conversation, switching topics, changing focus, deflection, irrelevant comments; Lack of depth, simple solutions offered to complex problems; Tone or language that is sarcastic / dismissive / aggressive / threatening / conspiratorial / derogatory; Personal attacks, ad hominem attacks, and personal criticism; Provocation; Misinformation or misrepresentation of facts; Speculative accusation

Characteristics of good faith comments: Acknowledgement of the issue; Engaged with / relevant to the original topic; Reasoned agreement or evidence-based disagreement; Constructive Argument / Criticism / Inquiry; Solution-oriented or strategic comments; Tone that is respectful / positive / playful / non-inflammatory; Encourages further interaction; Constructive speculation; Concern for accuracy

Now consider this tweet from <ACCOUNT>: <ORIGINAL TWEET TEXT>.
A user replies: <REPLY TEXT>

Using the criteria above, is the reply a good faith interaction with the original? Please answer in one word, yes or no".

Examples of good faith interaction include the following:

National Park Service: Living your life to the fullest does not have to involve selfies with bison.
Reply: Ya really should not have to tweet this, yet here we are

NBC News: Analysis by Philip Bump: Why would a news organization pay someone to lie to viewers?
Reply: Good question. I'm glad the journalists and NBC News hosts are speaking out against the hiring of Ronna McDaniel. She actually participated in the attempt to overthrow the free, fair & most secure election in the U.S. She belongs in jail, not normalized on NBC.

Examples of bad faith interactions are:

CBS News: JPMorgan Chase CEO Jamie Dimon says he worries geopolitical events and U.S. political polarization "may very well be creating risks that could eclipse anything since World War II."
Reply: Risks? Like illegal immigration, rigged elections, sideshows on freeways and violent criminals walking with no bail? Doesn't sound risky at all

NASA: LIVE: A new spacecraft is launching with crew for the first time! @NASA_Astronauts Butch Wilmore and Suni Williams are scheduled to lift off on @BoeingSpace 's #Starliner Crew Flight Test, riding aboard a @ULALaunch Atlas V rocket, at 12:25pm ET (1625 UTC).

Reply: So so old. Are they able to move around on their own? You better send a nurse just in case they forget what they are doing. now I see why the other countries mock us.

## 3 Results

### 3.1 Coding Accuracy Assessment

Of 400 tweet-reply pairs, 397 were coded (3 dropped for non-English content). Human coders agreed 87.7% of the time (Cohen's κ = 0.64). ChatGPT achieved 89.0% agreement with final human labels (κ = 0.75). For good faith detection, ChatGPT achieved 84.43% precision and 81.75% recall. For bad faith detection, precision was 91.64% and recall was 92.98%.

|  |  | Human | |
|---|---|---|---|
|  |  | Good | Bad |
| ChatGPT | Good | 103 | 19 |
|  | Bad | 23 | 252 |

**Table 1.** Confusion Matrix of ChatGPT vs. Human Coding for Good and Bad Faith Replies.

The moderate human inter-rater reliability (κ = 0.64) highlights the inherent challenge in assessing good and bad faith interactions, as noted in theoretical work [10][11]. ChatGPT's strong performance against human ground truth suggests LLMs can reliably identify these patterns.

### 3.2 Good and Bad Faith Interactions on Twitter/X

In our human-coded sample (N=397), 31.7% of replies were good faith. There were stark differences between account types: media accounts received 20.8% good faith replies versus government accounts' 39.7%.

Verification status showed significant patterns. Among unverified accounts (N=246), 37.8% of replies were good faith. For verified accounts (N=151), only 21.9% were good faith. ChatGPT labeling showed similar distributions with no significant differences (p=0.76 overall, p=0.67 for verified, p=0.92 for unverified).

Using ChatGPT to label all 31,283 replies revealed: 24.9% good faith overall, but only 18.7% from verified users versus 28.5% from unverified users. Verified accounts produced 81.3% bad faith interactions compared to 71.5% from unverified accounts.

### 3.3 Algorithmic Amplification Patterns

Verified status strongly influences reply ranking. The correlation between rank and percentage of verified account tweets at that rank is r=-0.85, indicating verified accounts dominate top-ranked positions. Average reply rank differs significantly (p<0.001): 32.8 for verified accounts versus 59.8 for unverified accounts.

While we found no significant correlation between rank and good faith percentage (r=0.18), the combination of verified accounts' higher bad faith rates and their algorithmic amplification raises concerns about discourse quality in highly visible reply positions.

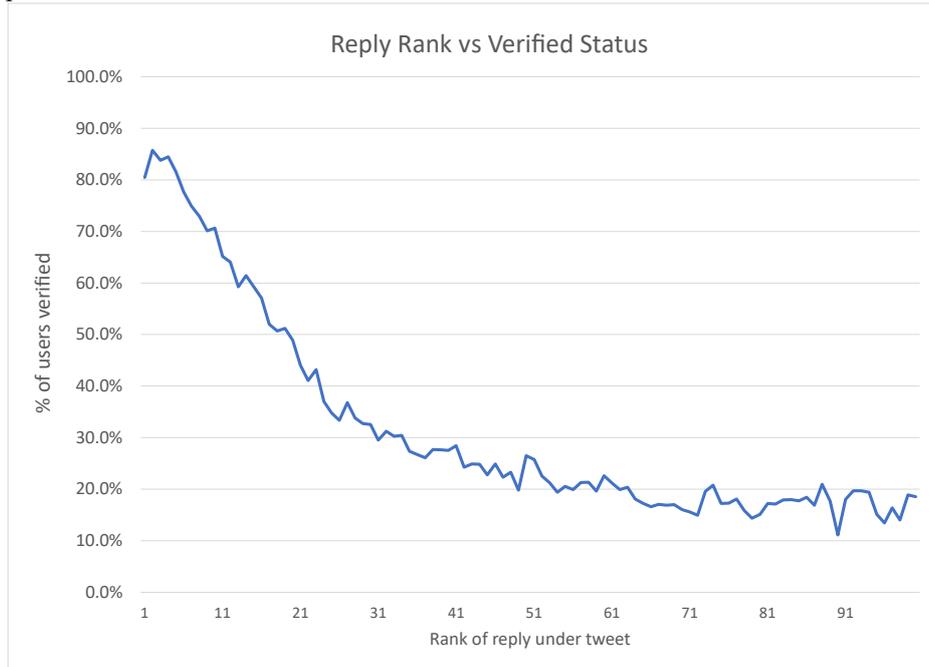

**Fig. 1.** Percentage of Verified Users by Reply Rank.

## 4 Discussion

### 4.1 Discussion Quality in High-Visibility Contexts

Within our specific sample of high-engagement posts from mainstream media and government accounts, bad faith interactions dominate (68.3% overall). This pattern is pronounced for verified accounts (81.3% bad faith), which are algorithmically amplified to prominent positions. While these findings cannot be generalized to all Twitter/X conversations, they suggest concerning patterns in how users engage with authoritative information sources that serve important democratic functions.

The high rate of bad faith interactions from verified accounts may reflect the parasocial nature of engagement on Twitter/X, where replies often function as performances for one's own followers rather than genuine dialogue attempts [13]. Strategic harassment of high-visibility targets can increase the attacker's own visibility within the platform ecosystem, creating perverse incentives for bad faith engagement.

### 4.2 Platform Design Implications

Our findings suggest Twitter/X's algorithmic amplification may inadvertently reward inflammatory content by prioritizing verified accounts that show higher rates of bad faith engagement. If platforms prioritized conversational health, they could consider reply quality as a ranking factor. Given that LLMs can identify good/bad faith replies with accuracy comparable to human coders, such interventions are technically feasible.

The tensions between platform business models and conversational health present significant challenges. Many social media platforms operate on engagement metrics that may conflict with discourse quality [14]. Interventions to improve conversational health must contend with these economic realities, potentially requiring metrics that better align platform success with user wellbeing rather than raw engagement numbers.

### 4.3 Limitations and Future Work

Our focus on direct replies to mainstream media and government accounts provides insight into engagement with authoritative sources but limits generalizability. The binary classification of good versus bad faith represents a necessary simplification of complex communicative behaviors. Future work should examine whether similar patterns appear in other conversational contexts and explore more nuanced categorization schemes.

The moderate human inter-rater reliability also suggests that determining faith can be subtle and context-dependent, indicating potential benefits from more detailed coding frameworks that capture different types or degrees of bad faith engagement.

## 5 Conclusion

We developed a framework for identifying good and bad faith interactions on social media and demonstrated that ChatGPT can label English-language replies with accuracy comparable to human coders. Within our studied sample of high-engagement posts from mainstream media and government accounts, 68.3% of replies were bad faith, with verified accounts showing particularly high rates (81.3%).

While these findings are specific to our sample context, they raise concerns about discourse quality around authoritative information sources. The combination of high bad faith rates from verified accounts and their algorithmic amplification suggests platform design choices may inadvertently undermine conversational health.

Future work should examine scalable approaches to improving online discourse quality while considering the complex relationships between platform economics, user behavior, and democratic engagement. By advancing understanding of good faith interactions and developing ethical ways to promote them, we can work toward creating more constructive digital public spaces.